\documentclass[aps, twocolumn, nofootinbib,superscriptaddress]{revtex4-1}
\usepackage{graphicx}
\usepackage{epsfig}
\usepackage{amsmath}
\usepackage{amsfonts}
\usepackage{amssymb}
\usepackage{hyperref}
\usepackage{tikz}
\usetikzlibrary{arrows}
\usetikzlibrary{snakes}

\newcommand{\ket}[1]{|{#1}\rangle}

\newcommand{\ketbra}[2]{|{#1}\rangle\langle{#2}|}

\newcommand{\trace}{{\rm Tr}}
\newcommand{\beq}{\begin{equation}}
\newcommand{\eeq}{\end{equation}}
\newcommand{\bea}{\begin{eqnarray}}
\newcommand{\eea}{\end{eqnarray}}

\begin{document}

\author{Cecilia Cormick} 
\affiliation{Departamento de F\'\i sica, FCEyN, UBA, \& IFIBA CONICET, 
Ciudad Universitaria Pabell\'on 1, 1428 Buenos Aires, Argentina} 
\author{Juan Pablo Paz} 
\affiliation{Departamento de F\'\i sica, FCEyN, UBA, \& IFIBA CONICET, 
Ciudad Universitaria Pabell\'on 1, 1428 Buenos Aires, Argentina} 

\title{Observing different phases for the dynamics of entanglement in an ion trap}


 
\date{\today}

\begin{abstract}
The evolution of the entanglement between two oscillators coupled to a common thermal environment is non-trivial. The long time limit has three qualitatively different behaviors (phases) depending on parameters such as the temperature of the bath ({\em Phys. Rev. Lett.} \textbf{100}, 220401). The phases include cases with non-vanishing long-term entanglement, others with a final disentangled state, and situations displaying an infinite sequence of events of disappearance and revival of entanglement. We describe an experiment to realize these different scenarios in an ion trap. The motional degrees of freedom of two ions are used to simulate the system while the coupling to an extra (central) ion, which is continuously laser cooled, is the gateway to a decohering reservoir. The scheme proposed allows for the observation and control of motional entanglement dynamics, and is an example of a class of simulations of quantum open systems in the non-Markovian regime.
\end{abstract}

\maketitle

\section{Introduction}

Entanglement is not only an essential feature of the quantum world but also a physical resource enabling the manipulation of information in non-classical ways \cite{Nielsen}. Understanding the evolution of entanglement for open systems is thus important from both a fundamental and a practical viewpoint. In fact, developing novel quantum information technologies requires a thorough characterization of the process of decoherence by which entanglement is typically degraded.

Considerable efforts were made in this direction and it was shown that even simple quantum open systems exhibit non-trivial features in the evolution of entanglement. For instance, it was observed that entanglement can vanish in finite time even if coherences only decay asymptotically \cite{Yu-Eberly-2004}. The long-time behaviour of entanglement for qubits interacting with bosonic baths has been analyzed in \cite{Scala-Migliore-Messina-2008}, where different asymptotic regimes where identified. Recently, a system formed by two resonant oscillators coupled to a common thermal reservoir was found to exhibit three different dynamical phases characterizing the asymptotic entanglement \cite{paz-roncaglia}. 

In this paper we propose and analyze an experiment to observe these three phases, using an array of three cold ions in a linear trap. This experiment can be viewed as a simple instance of a quantum simulation of the evolution of quantum open systems in the non-Markovian and non-perturbative regime. Trapped ions simulators were introduced in \cite{Wineland-et-al-1998} and are currently under active investigation \cite{Myatt-et-al-2000, Leibfried-et-al-2002, Porras-Cirac-2004, Schmidt-Kaler-2008, Friedenauer-2008}. Our proposal builds on these ideas in order to study the dynamics of entanglement between motional degrees of freedom. We note that motional entanglement has recently been demonstrated in ion traps in a static regime \cite{jost-2009}. 

Let us first describe the problem studied in \cite{paz-roncaglia} and then show how its essential features could be reproduced in an ion trap. The subsystems are two equal harmonic oscillators identically coupled to a common oscillator bath (the coupling is bilinear in position). In the long time limit there are three qualitatively different possible behaviors (phases), depending on parameters such as the bath temperature, the squeezing of the initial state, etc.: 1) The system may end up containing a non-vanishing amount of entanglement for all sufficiently long times, or 2) there may be an infinite sequence of events of disappearance followed by revivals of the entanglement, or 3) the asymptotic entanglement may be exactly zero. 

\begin{figure}
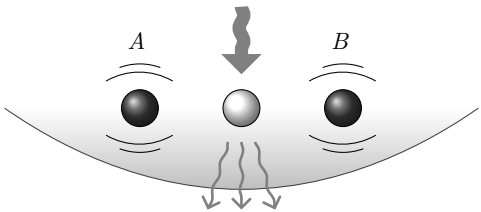
 
\begin{center}
\tikz[scale=0.9, transform shape, >= angle 90]{ 
\shade[top color=white,bottom color=lightgray] (-1.5,0) parabola bend (2,-1.2) (5.5,0); 
\draw[color=darkgray] (-1.5,0) parabola bend (2,-1.2) (5.5,0); 
\shadedraw[ball color=darkgray, draw=black] (0.5,0) circle (2ex); 
\shadedraw[ball color=white, draw=black] (2,0) circle (2ex);
\shadedraw[ball color=darkgray, draw=black] (3.5,0) circle (2ex);
\draw  
(0,0.4) arc (120:60:28pt) (0,-0.4) arc (-120:-60:28pt)
(0.2,0.6) arc (110:70:25pt) (0.2,-0.6) arc (-110:-70:25pt)
(3,0.4) arc (120:60:28pt) (3,-0.4) arc (-120:-60:28pt)
(3.2,0.6) arc (110:70:25pt) (3.2,-0.6) arc (-110:-70:25pt);
\draw[font=\itshape] (0.45, 1.0) node [scale=1.1]{A} (3.45,1.0) node [scale=1.1]{B};
\draw[snake=snake, segment amplitude = 0.9pt, line width=5pt, color=gray] (2,1.5) -- (2,0.7);
\fill[gray] (2,0.5)--(1.7,0.8)--(2.3,0.8)--cycle;
\draw[snake=snake, segment amplitude = 0.6pt, line width=1pt, color=gray, ->] (1.8,-0.5) -- (1.5,-1.5);  
\draw[snake=snake, segment amplitude = 0.6pt, line width=1pt, color=gray, ->] (2,-0.5) -- (2,-1.5);
\draw[snake=snake, segment amplitude = 0.6pt, line width=1pt, color=gray, ->] (2.2,-0.5) -- (2.5,-1.5);
}
\end{center}
\caption{A linear chain of three trapped ions. The ions at the ends will be equal and correspond to the subsystems. The central ion, coupled to a cooling laser, will be the source of decoherence of the system. \label{fig:three_ions}}
\end{figure}

These three possibilities exist in this simple system but have never been observed in the laboratory. Recently, a scheme has been proposed to observe two of these phases in a cavity-QED implementation \cite{Drumond-2009}. Our goal here is to propose an experiment in which by varying accesible parameters the three different behaviors may be attained, and detected. The general idea is illustrated in Fig \ref{fig:three_ions}: We consider a linear harmonic trap with three ions. The two ions at the ends of the chain are assumed to be equal and will constitute the subsystems, while the central ion will provide an environment with which the subsystems interact. In turn, the motion of this ion will be laser-cooled (pumping energy out of the normal modes in which this ion takes part). We will show that these ingredients are sufficient for the appearance of the three entanglement phases. Either the axial or transverse degrees of freedom can be considered; we analyze first the case of transverse modes, and mention later the differences that arise if one uses axial modes.

The paper is organized as follows: In section \ref{sec:model} we introduce our system and its dynamics. Section \ref{sec:asymptotic entanglement} is devoted to the asymptotic state and its entanglement properties. In section \ref{sec:preparation and phases} we consider a family of initial states and find the corresponding phase diagrams for the asymptotic entanglement. Section \ref{sec:detection} explains how the different behaviours can be detected. Finally, we conclude in section \ref{sec:conclusion}.

\section{Description of the system}\label{sec:model}

We consider a linear chain of three ions, neglecting the motion along one of the transverse directions ($y$) in which the trapping is assumed to be tight. The motional Hamiltonian, in terms of their longitudinal and transverse coordinates ($z$ and $x$ respectively), is \cite{Retzker-2008, Morigi-Fishman}: 
\begin{eqnarray}
        H &=& \sum_{j=1}^{3} \left[\frac{p_j^2}{2m_j} +
        \frac{m^2}{2m_j} \omega_x^2 x_j^2 +
        \frac{m}{2} \omega_z^2 z_j^2
        \right]\nonumber\\
        &&+ \frac{e^2}{4\pi\epsilon_0}\sum_{i > j}^3
        \frac{1}{\sqrt{( x_i- x_j)^2
        + (z_i- z_j)^2}}
        \label{Hamiltonbasic}
\end{eqnarray}
Here, $m$ is the mass of the ions at the ends of the chain, while $\omega_{z}$ and $\omega_x$ are the trapping frequencies for these ions in the axial and transverse directions respectively. The transverse trapping frequency scales inversely with the mass of the ion, so it will be different for the central ion, which will be of a different species \cite{Wineland-1998}. 
We choose frequencies so that the stable configuration is a linear array, and the harmonic approximation is valid. The equilibrium distance between ions is then given by:
\beq 
d^3_{eq} = \frac{5e^2}{(16\pi m\omega_z^2\epsilon_0)}
\eeq 
and the linearized Hamiltonian for the transverse motion is:
\beq
        H = \sum_j \frac{1}{2 m_j} p_j^2
        + \frac{1}{2}\sum_{ij} \gamma_{ij} x_i
        x_j. 
\eeq 
Here, \cite{Retzker-2008}
\bea
\gamma_{ii}&=&\frac{m^2}{m_i}\omega_x^2-\sum_j \frac{e^2}{2\pi\epsilon_0\vert z_i-z_j\vert^3}, \\
\gamma_{ij}&=&\frac{e^2}{2\pi\epsilon_0 \vert z_i-z_j\vert^3} \quad {\rm for}~ i\neq j. 
\eea 
Because of the symmetry, the normal modes of this Hamiltonian have well defined parity: there are two even modes, which involve the motion of all ions, and one odd mode in which the central ion is at rest. 

As mentioned before, we associate the two ions at the ends of the chain with two susbystems $A$ and $B$. The central ion provides an effective environment, and the two subsystems are equally coupled to it. By modifying the ratio $\omega_x/\omega_z$, the strength of the system-environment coupling is modified (when $\omega_x\gg\omega_z$ the normal modes are approximately equal to the local modes so the coupling tends to vanish in this limit). 

The use of collective coordinates $x_\pm = (x_A\pm x_B)/\sqrt{2}$ is convenient since the center of mass of the system, $x_+$, is coupled to the environment while the relative motion, $x_-$, is not. In this way the model is analogous to the one studied in \cite{paz-roncaglia}, the main difference being that the environment consists of a single oscillator. 
However, the effective size of the environment can be enlarged by coupling the central ion with a laser tuned to cool down the modes in which this ion participates. This produces an effect similar to that of a low-temperature reservoir.
 
A useful approximation for the effect of the laser coupling can be obtained in interaction picture and assuming that the system is in the Lamb-Dicke regime. Then, if the laser is tuned to the red sideband of one of the motional modes ($n$, where $n$ labels either the center of mass or the third mode), after a rotating wave approximation we find the Hamiltonian \cite{Leibfried-et-al-2003}:
\beq
H_{L-I}^{(int)} = i \frac{\hbar\Omega_R}{2} v^{(n)}_2 k\sqrt{\frac{\hbar}{2m\nu_n}} (\sigma_+ a_n e^{i\varphi} - h.c.)
\eeq
Here $\Omega_R$ is the Rabi frequency, $k$ is the component of the wave vector along the mode direction, and $\varphi$ is the phase of the laser. $\sigma_\pm$ are the transition operators for the internal state of the ion, $\sigma_+ = \ketbra{e}{g}$, $\sigma_- = \ketbra{g}{e}$, while $a_n, a_n^\dagger$ are annihilation and creation operators for the normal mode. 

Through this coupling, excitations can be transferred between the motional modes and the internal state of the ion. We shall assume that $\ket{g}$ is a stable internal state, while $\ket{e}$ decays back to $\ket{g}$. Thus, the effect of the coupling to the laser will be to cool down the motional mode $n$ \cite{Eschner-et-al-2003}. We shall assume that each of the modes in which the central ion participates is damped in this way. The setup we are considering is thus a particular case among the ones studied in \cite{Kielpinski-et-al-2000} for sympathetic cooling.

\section{Asymptotic entanglement}\label{sec:asymptotic entanglement}

As in \cite{paz-roncaglia}, we shall restrict our analysis to Gaussian states. A Gaussian state $\rho$ is fully characterized by the values of the first moments, $\trace(\rho R_j)$, and the covariance matrix:
\bea
C_{jk} &=& \frac{1}{2} \trace(\rho \{R_j, R_k\}) - \trace(\rho R_j) \trace(\rho R_k ), \\
R &=& (x_1, x_2, x_3, p_1, p_2, p_3)
\eea
where the curly brackets denote the anticommutator. The entanglement properties are only determined by the covariance matrix $C$, since changes in the first moments can always be achieved by local operations. 

According to our simple model for laser cooling, the odd mode of the chain, corresponding to coordinates $x_-$, $p_-$, evolves following its own free dynamics. In contrast, the even modes approach the vacuum. Thus, the second moments of $x_+$, $p_+$ approach equilibrium, characterized by $\langle \{x_+,p_+\} \rangle=0$ and:
\bea
\langle x_+^2\rangle ~=~ \Delta^2x_+ &=& \frac{\hbar}{2m} \left(\frac{c_{e1}^2}{\omega_{e1}}  + \frac{c_{e2}^2}{\omega_{e2}}  \right),\\
\langle p_+^2\rangle ~=~ \Delta^2p_+ &=& \frac{m\hbar}{2} \left(\omega_{e1}~ c_{e1}^2 + \omega_{e1}~ c_{e2}^2 \right).
\eea
Here, $e1$ and $e2$ label the two even modes, $\omega_{e1/e2}$ are the  corresponding frequencies, and $c_{e1/e2}$  the amplitude coefficients transforming $x_+$ to the even normal modes. 

From the covariance matrix we can calculate the entanglement between $A$ and $B$, quantified by the logarithmic negativity $E_N$ \cite{serafini-et-al-2004}. If the odd mode is squeezed, the entanglement between subsystems may oscillate as this mode evolves. The logarithmic negativity reaches its maximum and minimum values when the dispersions $\Delta x_-$, $\Delta p_-$ are extremal (\emph{i.e.} when the odd mode squeezing is along the position or momentum axis) and at these times it is given by \cite{paz-roncaglia}: 
\beq \label{eq:negativity}
E_N=\max\left\{0, -\ln\left(\frac{2\Delta p_- \Delta x_+}{\hbar}\right),  -\ln\left(\frac{2\Delta x_- \Delta p_+}{\hbar}\right)\right\}.
\eeq
More generally, the asymptotic behaviour of entanglement is determined by the following three quantities characterizing the final state \cite{paz-roncaglia}:
\bea
r &=& \frac{1}{2} \left|\ln\left(m\omega_{odd}\frac{\Delta x_-}{\Delta p_-}\right)\right|,\\
r_{crit} &=& \frac{1}{2} \left|\ln\left(m\omega_{odd}\frac{\Delta x_+}{\Delta p_+}\right)\right|,\\
S_{crit} &=& \frac{1}{2} \ln\left(\frac{4}{\hbar^2}\Delta x_+ \Delta p_+ \Delta x_- \Delta p_-\right),
\eea
where $\omega_{odd}$ is the frequency of the odd mode. Here, $r$ is the squeezing of the odd mode, $r_{crit}$ is related to the asymptotic squeezing of the system's center of mass, and $S_{crit}$ to the final entropy of the system formed by the ions $A$ and $B$. These parameters depend on the choice of ions and the ratio between the axial and transverse frequencies, as well as on the initial state of the odd mode. 

\begin{figure}[h,b,t]
\begin{center}
    \epsfig{file=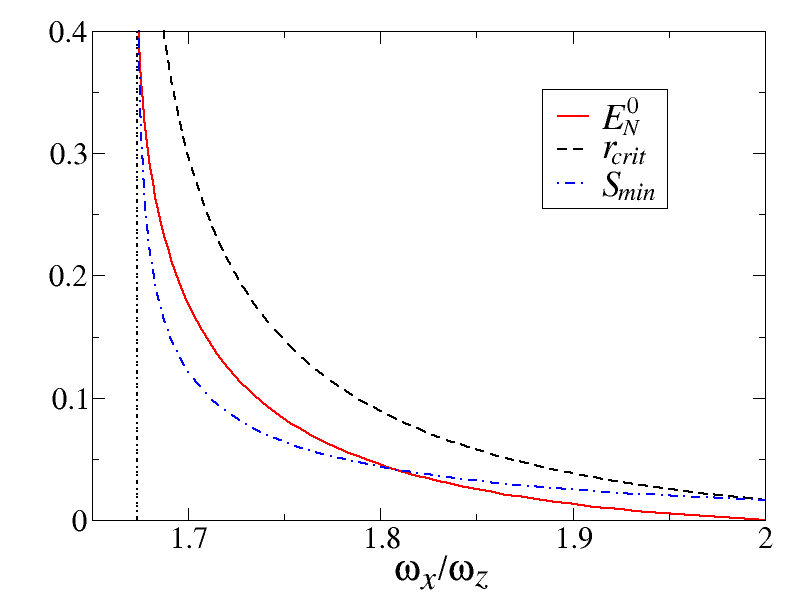, angle=0, width=0.45\textwidth}
\end{center}
\vspace{-15pt} 
\caption{Values of the parameters $E_N^{0}$ (red, continuous), $r_{crit}$ (black, dashed), and $S_{min}$ (blue, dash-dotted) as functions of the ratio $\omega_x/\omega_z$ for a chain of two $^{24}$Mg ions with one $^9$Be ion placed between them. For  values of $\omega_x/\omega_z$ smaller than 1.673\dots, as indicated by the vertical line, the linear configuration becomes unstable.}
\label{fig:parametros}
\end{figure}

The possibility to observe the three dynamical regimes in our model is conditioned by the values that these three parameters may take \cite{paz-roncaglia}. Typically, entanglement will be large for large squeezings, and will decrease with increasing entropy. For instance, the entanglement between subsystems when the chain is in its vacuum state is given by the difference:
\beq
E_N^{0} = \max\{r_{crit} - S_{min}, 0\},
\eeq
with 
\beq
S_{min} = \frac{1}{2} \ln\left(\frac{2\Delta x_+ \Delta p_+}{\hbar}\right)
\eeq
the value for the entropy of the system formed by ions A and B when the chain is in its ground state (in general $S_{crit}\geq S_{min}$, for a given choice of masses and frequencies). In order to obtain large values for $r_{crit}$, we need the coupling between ions to be strong, which in turn implies being not too far away from the transition between linear and zig-zag equilibrium configurations \cite{Morigi-Fishman}. On the other hand, the parameters must be chosen in such a way that the harmonic approximation is valid (\textit{i.e.} not too close to the  transition). To achieve good values for $E_N^{0}$, it turns out convenient to have a central ion lighter than the other two. By using different ion species for the subsystems and the environment one also reduces the unwanted coupling of the laser to the subsystem ions. For the case in which $A$, $B$ are $^{24}$Mg ions and the central ion is $^9$Be, Fig. \ref{fig:parametros} displays the values of $r_{crit}$, $S_{min}$ and $E_N^{0}$ as functions of $\omega_x/\omega_z$. We note that these species have been used to demonstrate motional entanglement in \cite{jost-2009}.  

\section{State preparation and entanglement phases}\label{sec:preparation and phases}

In order to make the different asymptotic situations actually observable one should find a procedure to prepare a family of suitable initial states. Conceptually the simplest initial state is the vacuum of the chain Hamiltonian which, as pointed out in \cite{retzker-et-al-2005}, is a squeezed entangled state in terms of the local oscillators $A$ and $B$. In this case, the system does not evolve in time, and the entanglement is given by $E_N^0$. 
More generally, one can start with a thermal state instead of the vacuum. This also corresponds to a Gaussian state and leads to a situation with a larger final entropy, $S_{crit}> S_{min}$, without changing the values of $r$ and $r_{crit}$. Actually, the asymptotic state depends on the initial temperature for the odd mode only. By varying this parameter, associated to the mean population $\langle n_-\rangle$ of the mode, the final entropy is modified. Thermal states were experimentally obtained in \cite{Meekhof-1996} from Doppler cooling of a single trapped ion.

The squeezing of the initial state can in turn be varied by an abrupt relaxation of the transverse trapping frequency, $\omega_x' \to \omega_x = \omega_x'/f$. After this operation the dispersions remain untouched, and since the frequency $\omega_{odd}$ changes, the squeezing of the odd mode is modified to $r = \frac{1}{2} \ln(f)$. $S_{crit}$ and $r_{crit}$ depend on the final value of $\omega_x$ but not on the expansion factor $f$. More sophisticated ideas concerning the creation of squeezing using time modulation of the trapping potentials were discussed in \cite{Serafini-Retzker-Plenio-2009}. An alternative method was presented in \cite{Meekhof-1996}, where the squeezing in the motion of a single trapped ion was experimentally induced obtaining values $r\approx2$. This was achieved by the coupling to a laser with a detuning equal to twice the frequency of the mode, and the same procedure could be used here to squeeze the odd mode. 

\begin{figure}[h,b,t]
\begin{center}
    \epsfig{file=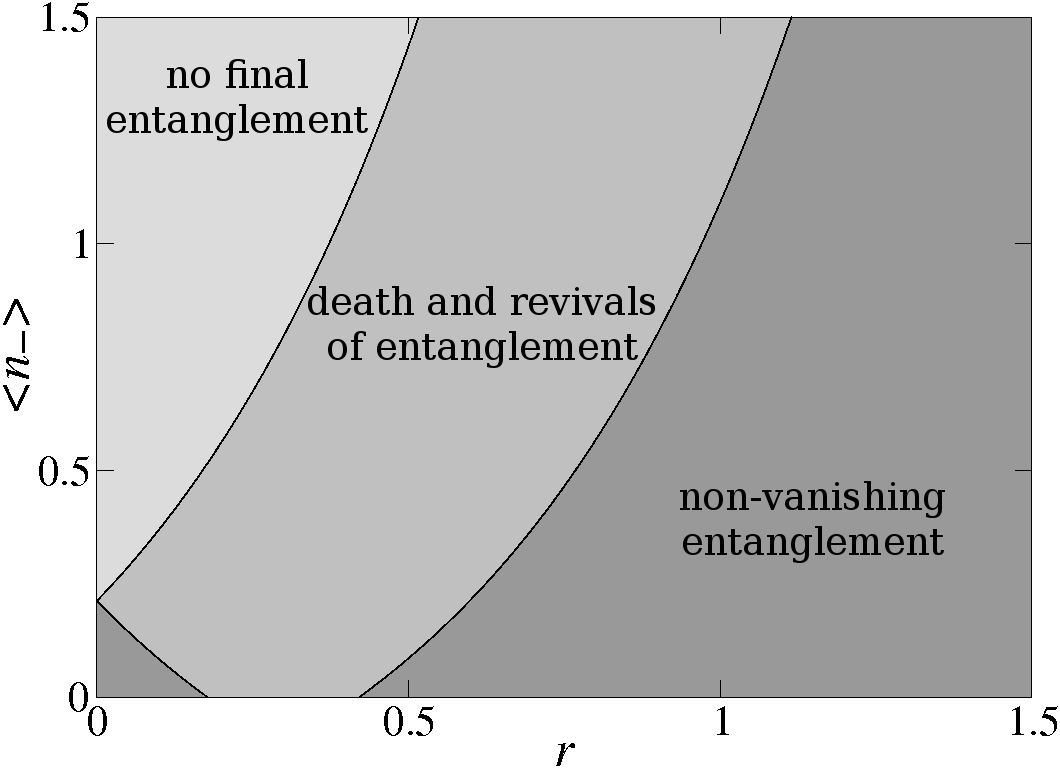, angle=0, width=0.45\textwidth}
\end{center}
\vspace{-15pt}
\caption{Asymptotic behaviour for the entanglement in a Gaussian state parameterized by the initial mean population of the odd mode, $\langle n_-\rangle$, and the subsequently induced squeezing of the mode, $r$. The width of the area corresponding to the phase with death and revivals is determined by the value of $r_{crit}$, which depends on the choice of ions and on the ratio $\omega_x/\omega_z$. The case plotted corresponds to a chain of three ions, one $^9$Be ion in the middle and two $^{24}$Mg ions at the ends; the trapping frequencies are chosen in the form $\omega_x=1.7 ~\omega_z$. 
} \label{fig:phase_diagram}
\end{figure}

\begin{figure}[h,b,t]
\begin{center}
    \epsfig{file=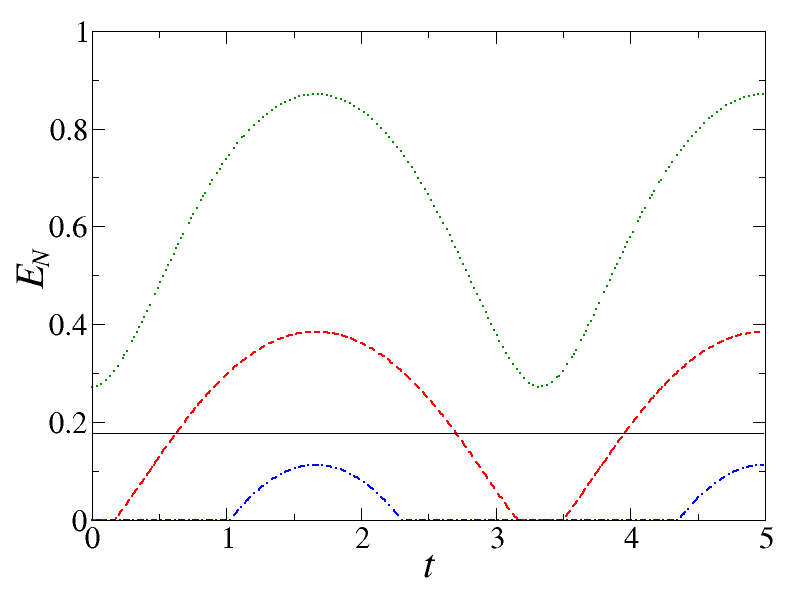, angle=0, width=0.45\textwidth}
\end{center}
\vspace{-15pt}
\caption{Different asymptotic behaviours for entanglement (measured by the logarithmic negativity) as a function of time. The choice of ions and frequencies is the same as in the previous figure. The cases plotted correspond to different values of the parameters $\langle n_-\rangle$ for the initial population of the odd mode and $r$, the subsequently induced squeezing: $\langle n_-\rangle = 0$, $r=0$ (black, continuous); $\langle n_-\rangle = 0$, $r=0.3$ (red, dashed); $\langle n_-\rangle = 0$, $r=1$ (green, dotted) and $\langle n_-\rangle = 1$, $r=0.7$ (blue, dash-dotted). Time is in units of $\omega_z^{-1}$, and the time origin is arbitrary. We note that the entanglement of a Bell pair would correspond to $E_N\simeq1$.}
\label{fig:negativity}
\end{figure}

One can in this way modify the two relevant parameters $S_{crit}$ and $r$, while the value of $r_{crit}$ is only determined by the transformation to the normal modes. It is thus possible to sample over the three asymptotic regimes, as shown in Fig. \ref{fig:phase_diagram} (for the case $\omega_x=1.7 ~\omega_z$). 
The behaviour of the asymptotic entanglement as a function of time is illustrated in Fig. \ref{fig:negativity} for different initial states.

\begin{figure}[h,b,t]
\begin{center}
    \epsfig{file=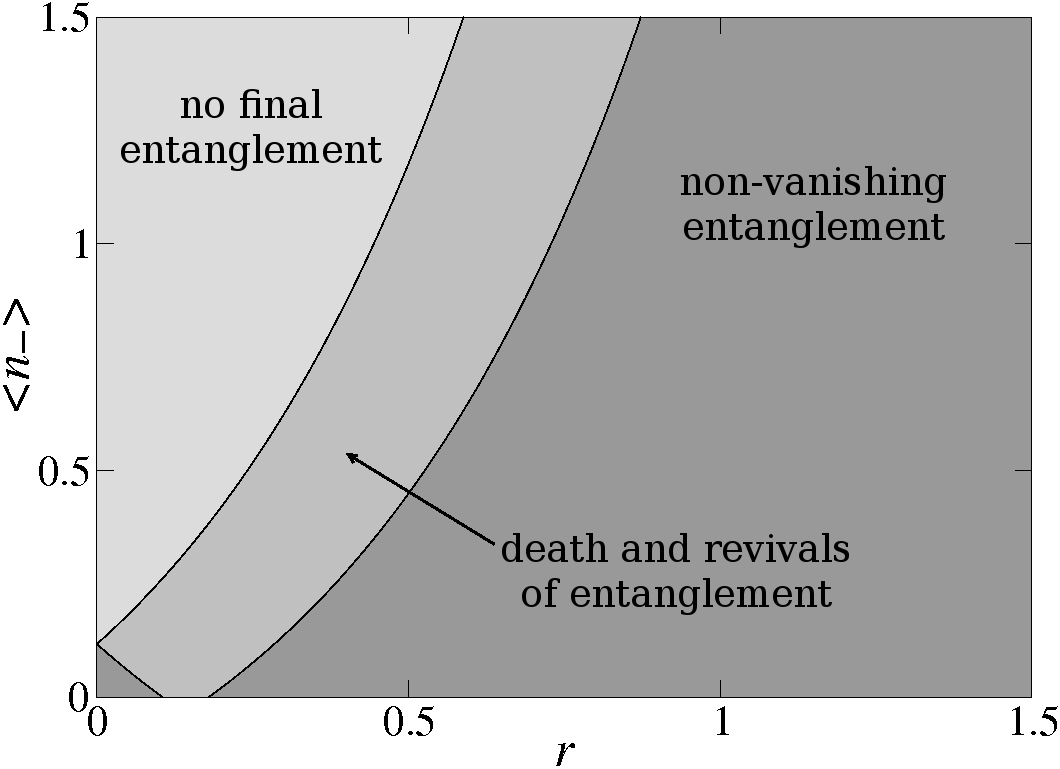, angle=0, width=0.45\textwidth}
\end{center}
\vspace{-15pt}
\caption{Asymptotic behaviour for the entanglement in a Gaussian state parameterized by the initial mean population of the odd mode, $\langle n_-\rangle$, and the subsequently induced squeezing of the mode, $r$, when the axial modes are considered. The case plotted corresponds to a chain of three ions, one $^9$Be ion in the middle and two $^{24}$Mg ions at the ends. 
} \label{fig:phase_diagram_axial}
\end{figure}

The axial motion of trapped ions could be used with similar results, though the typical values for $r_{crit}$ (and accordingly $E_N^0$) are smaller. For our choice of ions, the relevant parameters for the axial case are given by $r_{crit}\simeq0.14$, $S_{min}\simeq0.04$, compared to $r_{crit}\simeq0.30$, $S_{min}\simeq0.12$ for the transverse motion with $\omega_x=1.7 ~\omega_z$ as in Figs. \ref{fig:phase_diagram} and \ref{fig:negativity}. When the axial modes are considered, changing the value of the mass ratio between ions does not alter the results much: the phase diagram for our choice of ions, shown in Fig. \ref{fig:phase_diagram_axial}, is very similar to the one obtained for three equal ions. We note that in this case the technique of \cite{Meekhof-1996} seems to be the right strategy to generate squeezing, as the sudden alteration of trapping frequency would modify the equilibrium positions in such a way that the harmonic approximation could only hold for very small values of squeezing.

\section{Detection}\label{sec:detection}

For the experimental observation of the three entanglement phases, it is enough to measure the asymptotic dispersions $\Delta x_+$, $\Delta p_+$, and the time-varying quantities $\Delta x_-$, $\Delta p_-$. The discrimination of the different phases only requires the evaluation of the negativity (\ref{eq:negativity}) for the times when the dispersions take extremal values. The necessary variances can be inferred from $\Delta Q_n$, $\Delta P_n$, with $Q_n = (a_n+a_n^\dagger)/\sqrt{2}$, $P_n = i(a_n-a_n^\dagger)/\sqrt{2}$ the dimensionless coordinates for the normal modes. 

From blue-sideband oscillations together with fluorescence measurements, it is possible to reconstruct the population of the different levels for each mode \cite{Wineland-1998}. Thus, when the first moments vanish the covariance matrix $C$ for each mode can be determined in the following way: Firstly, the mean phonon number is related to the trace in the form: 
\beq
\trace(C) = 2 \langle a^\dagger a \rangle + 1.
\eeq 
Then, let us assume that we apply a laser pulse inducing a force which changes momentum in a value $\delta P$. The probability to find the mode in the ground state afterwards is given by \cite{Scutaru-1998}:
\beq
F(\delta P) = \sqrt{\det(M)} ~e^{-(\delta P)^2 M_{22}/2}, 
\eeq
with 
\beq
\quad M^{-1} = \frac{\mathbb{I}}{2}+C.
\eeq
From this set of measurements one can find $C$ (up to an unimportant sign of the off-diagonal elements) for each mode at each given time. In this protocol, only the force needs to be faster than the chain dynamics, since the populations are not modified by the free evolution. We note that the ability to perform measurements of time-dependent motional observables has recently been demostrated in \cite{Gerritsma-et-al-2009}, allowing for the quantum simulation of the Dirac equation.

 
\section{Concluding remarks}\label{sec:conclusion}

In conclusion, we have shown how the symmetry and interactions in the three-ion chain allow
us to reproduce the essential features of the system studied in \cite{paz-roncaglia}, including the three different entanglement phases. Furthermore, the observation of the asymptotic entanglement dynamics seems possible with available ion-trap techniques. As a last remark, we note that a more detailed investigation of the dynamics of the laser cooling process would allow for the description of the approach to the asymptotic regime, in terms of a non-Markovian master equation. The non-Markovian character of the process can be varied depending on the effective temperature of the bath and strength of the coupling between ions. The latter can be modified by the choice of trapping potential, while the former depends on the laser detuning. This would allow for the extension of the ideas in \cite{Poyatos-Cirac-Zoller-1996} for the case of more than one particle, and will be the subject of future research.

\end{document}